\documentclass[]{spie}  

\usepackage{amsmath,amsfonts,amssymb}
\usepackage{graphicx}
\usepackage[colorlinks=true, allcolors=blue]{hyperref}
\usepackage{underscore}
\usepackage{makecell}
\usepackage{bbold}

\title{BDG-Net: Boundary Distribution Guided Network for Accurate Polyp Segmentation}

\author{Zihuan Qiu\thanks{$\quad$First Author and Second Author contribute equally to this work.}}
\author{Zhichuan Wang$^*$}
\author{Miaomiao Zhang}
\author{Ziyong Xu}
\author{Jie Fan}
\author{Linfeng Xu\thanks{$\quad$Corresponding Author.}}
\affil{
University of Electronic Science and Technology of China, Chengdu, China
\\
lfxu@uestc.edu.cn
}

\begin{document} 
\maketitle

\begin{figure} [ht]
\begin{center}
\begin{tabular}{c} 
\includegraphics[height=4.5cm]{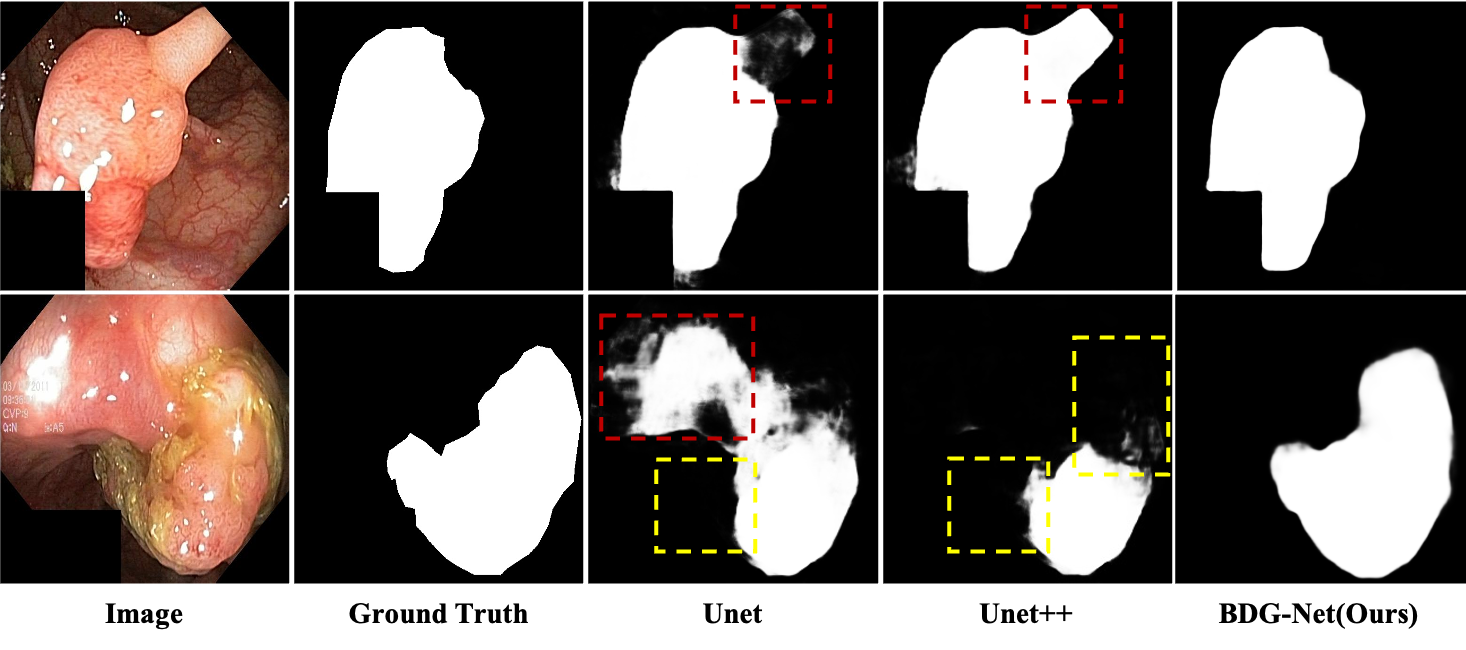}
\end{tabular}
\end{center}
\caption[example]
{ \label{fig1} 
Examples of polyp segmentation in the Kvasir\cite{jha2020kvasir} dataset. Red dashed box and yellow dashed box indicate over-segmentation and under-segmentation respectively. (Best viewed in color)}
\end{figure} 

\begin{abstract}
Colorectal cancer (CRC) is one of the most common fatal cancer in the world. Polypectomy can effectively interrupt the progression of adenoma to adenocarcinoma. Colonoscopy is the primary method to find colonic polyps. However, due to the different sizes and the unclear boundary of polyps, it is challenging to segment polyps accurately. To this end, we design a Boundary Distribution Guided Network (BDG-Net) for accurate polyp segmentation. Specifically, Boundary Distribution Generate Module (BDGM) aggregates high-level features to generate Boundary Distribution Map (BDM), which is sent to the Boundary Distribution Guided Decoder (BDGD) as complementary spatial information to guide the polyp segmentation. Moreover, a multi-scale feature interaction strategy is adopted in BDGD to improve the polyps segmentation of different sizes. Extensive experiments demonstrate that BDG-Net outperforms state-of-the-art models remarkably and maintains low computational complexity. Code: \url{https://github.com/zihuanqiu/BDG-Net}
\end{abstract}


\section{INTRODUCTION}
\label{sec:intro}  

Medical image segmentation is an essential part of the artificial intelligence-assisted diagnosis. It can provide some fine-grained information to assist diagnosis, such as the location and shape of the polyps. This information is critical and helpful for successful treatment, even preventing disease. Take colorectal cancer(CRC) as an example. CRC is the third most common type of cancer around the world\cite{silva2014toward}. Early detection through colonoscopy has been shown to be effective in reducing disease-related mortality.

Recently, the convolutional neural network (CNN) based medical image segmentation methods have achieved favorable performance in many datasets. Most of them are based on encoder-decoder structure. The most representative method is U-Net\cite{ronneberger2015u}, which captures precise context with skip paths. Oktay et al.\cite{oktay2018attention} introduced attention-mechanism to the segmentation network based on U-Net. Zhou et al.\cite{zhou2018unet++} proposed a variant of U-Net named UNet++, in which decoder sub-networks are connected by dense skip paths. PraNet\cite{fan2020pranet} proposed reverse attention to achieve accurate polyp segmentation.  However, due to the different sizes of polyps and the unclear boundary between polyps and their surrounding mucosa, segmentation models often suffer under-segmentation or over-segmentation, as shown in Figure \ref{fig1}. 

In this paper, we propose a Boundary Distribution Guided Segmentation Network (BDG-Net), which consists of two novel modules. The first one is the Boundary Distribution Generate Module (BDGM), aggregating the high-level features and generating Boundary Distribution Map (BDM) under the supervision of the ideal BDM. The second one is the Boundary Distribution Guided Decoder (BDGD), which uses the generated BDM as complementary information, fusing different multi-scale features to improve polyp segmentation on different sizes. Quantitative and qualitative experiments show that our proposed model outperforms state-of-the-art methods on five challenging polyp datasets while maintaining low computational complexity. 

The contributions of this work are summarized as follows: 
\begin{itemize}
    \item[(1)] We design BDGM to generate BDM, which can be used as complementary spatial information to guide the accurate polyp segmentation.
    \item[(2)] We propose BDGD, which fuses multi-scales features to improve polyp segmentation on different sizes.
    \item[(3)] Integrating the above two modules into a single architecture, we design BDG-Net, which outperforms state-of-the-art models remarkably on five challenging polyp datasets while maintaining low computational complexity.
\end{itemize}

\section{Methodology}
In this section, we ﬁrst give the definition of our proposed BDM and then introduce our two modules, BDGM and BDGD, respectively. Next, we specify the loss function we used. Finally, the entire architecture of the proposed BDG-Net is described.

\subsection{Boundary Distribution Map}
\label{sec:BDM}

Due to the unclear or illegible boundary of the polyps, it is challenging to clarify the polyps boundary accurately. However, it is much easier to estimate the range of the boundary. Based on this intuition, we propose BDM, which is a map representing the estimated probability that the current pixel belongs to the boundary. We assume that the boundary distribution follows a Gaussian distribution with a mean of 0 and a standard deviation of $\sigma$. The ideal BDM can be defined as follows:

\begin{equation}
BDM\left( p_{ij}\right)\  =\  \frac{1}{\sqrt{2\pi } \sigma } e^{-\frac{\varepsilon (p_{ij})^{2} }{2\sigma^{2} } }
\label{eq1}
\end{equation}

\noindent where $\varepsilon(p_{ij})$ is the shortest Euclidean distance from pixel $p_{ij}$ to the boundary and $\sigma$ is the standard deviation.


\begin{figure} [ht]
\begin{center}
\begin{tabular}{c} 
\includegraphics[height=7.5cm]{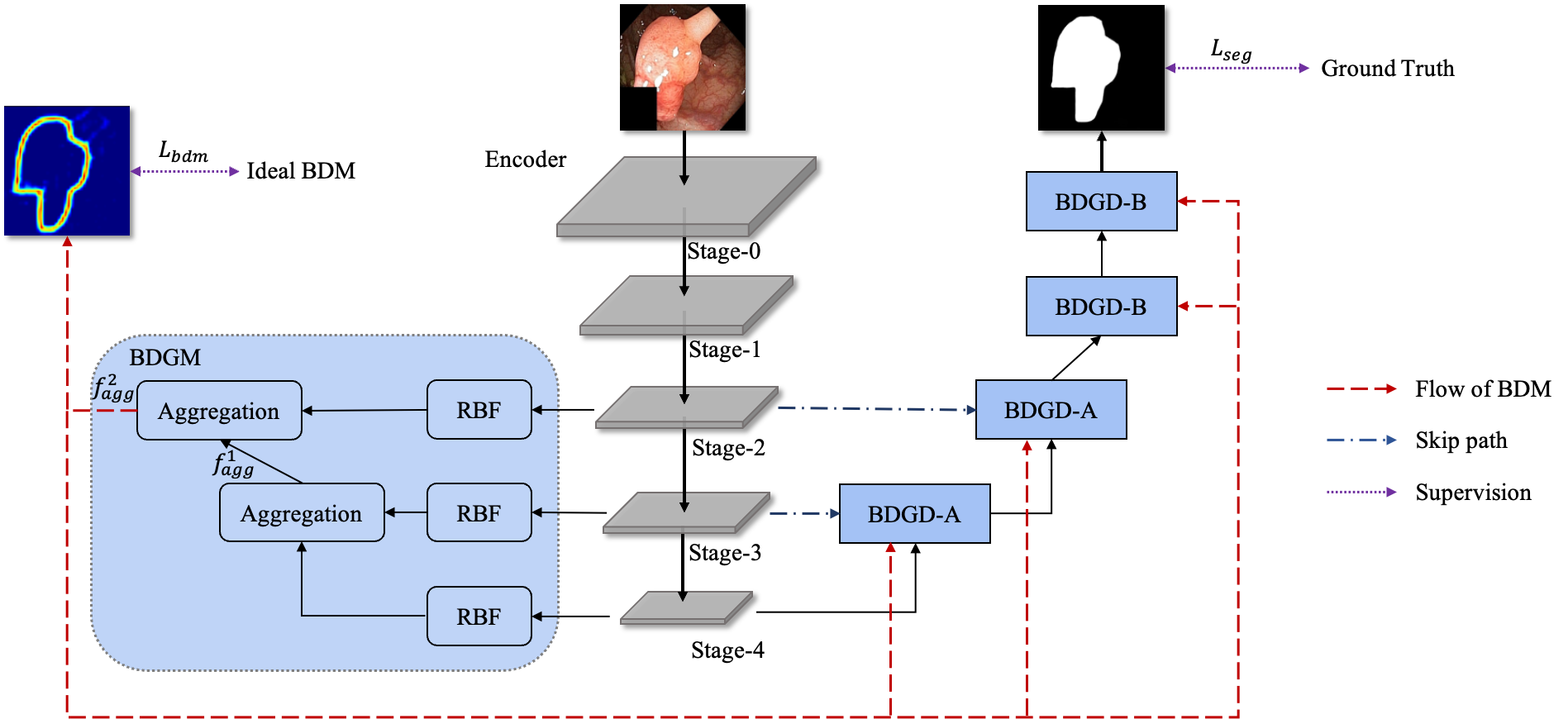}
\end{tabular}
\end{center}
\caption[example]
{ \label{fig2} 
Overall architecture of BDG-Net.}
\end{figure}

\subsection{Boundary Distribution Generate Module}

To predict the boundary distribution, we propose BDGM, which generates BDM under the supervision of the ideal BDM. BDGM aggregates the features from the last three stages of the encoder and gradually increases the feature resolution. As shown in Figure \ref{fig2}, we use the Receptive Field Block\cite{liu2018receptive}(RBF) to reduce the channel dimension of features and then aggregate the third and fourth-stage of encoder features to output $f_{agg}^{1}$, with resolution being ($\frac{H}{16}\times\frac{W}{16}$). Then the obtained feature $f_{agg}^{1}$ is aggregated with the second-stage features, and the resolution of the output feature $f_{agg}^{2}$ is ($\frac{H}{8}\times\frac{W}{8}$). Finally, we use bilinear interpolation to upsample by 8 to generate BDM.

\begin{figure} [ht]
\begin{center}
\begin{tabular}{c} 
\includegraphics[height=4.2cm]{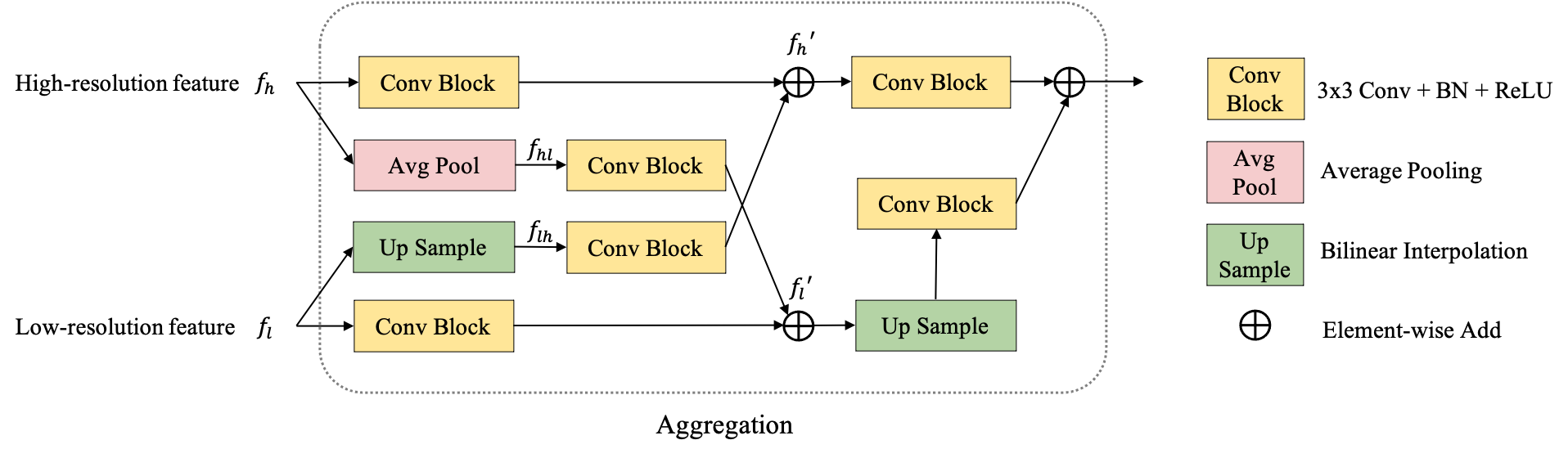}
\end{tabular}
\end{center}
\caption[example]
{ \label{fig4} 
Details of aggregation operation.}
\end{figure}

The aggregation is designed to maintain lightweight and eﬃcient. As shown in Figure \ref{fig4}, the high-resolution feature $f_h$ is downsampled by average pooling to get $f_{hl}$, and the low-resolution feature $f_l$ is upsampled by bilinear interpolation to get $f_{lh}$. Then $f_{lh}$ and $f_{hl}$ go through 3$\times$3 convolution followed by BN and ReLU to get $Conv(f_{lh})$ and $Conv(f_{hl})$, respectively. The convoluted $f_h$ is added with $Conv(f_{lh})$ to obtain feature $f'_h$. Similarly, The convoluted $f_l$ is added to $Conv(f_{hl})$ to obtain feature $f'_l$. Finally, the upsampled $f'_l$ goes through convolution block and adds with the convoluted $f'_h$ to get the output feature. The aggregation operation can be denoted as follows:
\begin{align}
    &f^{\prime }_{h}=Conv\left( f_{h}\right)  +Conv\left(f_{lh} \right)
    \label{eq2}  \\[0.1cm]
    &f^{\prime }_{l}=Conv\left( f_{l}\right)  +Conv\left( f_{hl}  \right)
    \label{eq3}\\[0.1cm]  
    &f_{out}=Conv\left( f^{\prime }_{h}\right)  +Conv\left( Up\left( f^{\prime }_{l}\right)  \right)  
    \label{eq4}
\end{align}
\noindent where $Conv(\dots)$ refers to convolutional layer with kernel size 3 followed by BN and ReLU. $Up(\dots)$ represents the upsampling by bilinear interpolation.

\subsection{Boundary Distribution Guided Decoder}

\begin{figure} [ht]
\begin{center}
\begin{tabular}{c} 
\includegraphics[height=8.0cm]{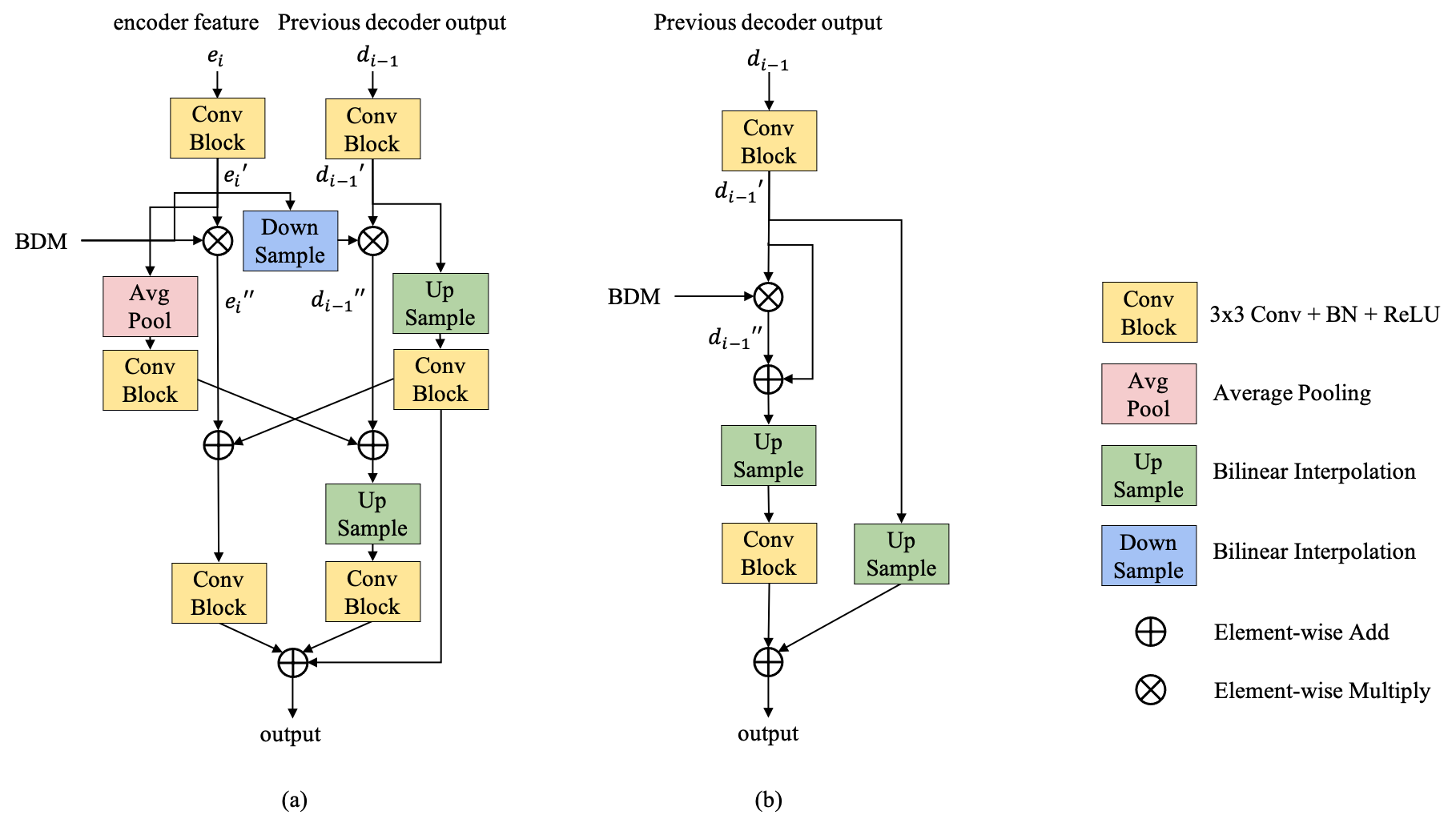}
\end{tabular}
\end{center}
\caption[example]
{ \label{fig5} 
Boundary Distribution Guided Decoder.}
\end{figure} 

BDGD uses the BDM generated by BDGM to guide the model segment accurately. In the higher stages, we use BDGD-A, which takes the output of the previous decoder and the encoder features as input, obtaining sufficient spatial information by interacting with the features on two scales. Specifically, as shown in Figure \ref{fig5}(a), let $e_i$ be the encoder feature and $d_{i-1}$ be the output of the previous decoder. Note that the resolution of $e_i$ is two times larger than $d_{i-1}$. Firstly, we transform $e_i$ and $d_{i-1}$ into output's channel dimensions through convolutional layer and get $e'_i$ and $d'_{i-1}$. Then, $e'_i$ is multiplied by BDM, and $d'_{i-1}$ is multiplied by the downsampled BDM, after which the boundary enhanced features $e''_i$ and $d''_{i-1}$ are obtained. The upsampled $d'_{i-1}$ goes through a ConvBlock and adds with $e''_i$ to get the high-resolution fused feature. The downsampled $e'_i$ goes through a ConvBlock and adds with $d''_{i-1}$ to get the low-resolution fused feature. Then the low-resolution fused feature is upsampled and added with the high-resolution fused feature to achieve cross-scale feature fusion. Finally, a residual structure is used to help optimization.

Similarly, in the lower stages, we adopt BDGD-B. The difference between BDGD-B and BDGD-A is that the branch to fuse encoder feature is removed, as shown in Figure \ref{fig5}(b). Experiments show that removing this branch does not affect the performance of segmentation while reducing the number of parameters and computational complexity.

\subsection{Loss function}
The loss function consists of $L_{bdm}$ and $L_{seg}$. $L_{bdm}$ is used to supervise the BDM, and $L_{seg}$ is used to supervise the final segmentation. $L_{bdm}$ optimizes the regions with higher loss and ignores the others. It conduces to the optimization of hard examples and improves the generalization ability of the model. $L_{seg}$ is the combination of weighted BCE loss ($L_{wbce}$) and weighted IoU loss ($L_{wiou}$). The total loss is as follows:
\begin{align}
    L_{total}&=L_{bdm} + L_{seg}=L_{bdm} + L_{wbce} + L_{wiou}
    \label{eq5}\\[0.1cm]
    L_{bdm}&=\sum\nolimits_{(i,j)} (b_{ij}-\hat{b}_{ij} )^{2}\cdot \mathbb{1} \left[ (b_{ij}-\hat{b}_{ij} )^{2}>\lambda \right]    
    \label{eq6}\\[0.1cm]
    L_{wbce}&=-\sum\nolimits_{\left( i,j\right)  } w_{ij}\cdot \left[ \hat{s}_{ij} log(s_{ij})+(1-\hat{s}_{ij} )log(1-s_{ij})\right] 
    \label{eq7}\\[0.1cm]
    L_{wiou}&=1-\frac{\sum\nolimits_{\left( i,j\right)  } [w_{ij}\cdot \hat{s}_{ij} \cdot s_{ij}]}{\sum\nolimits_{(i,j)} \left[ w_{ij}\cdot \left( \hat{s}_{ij} +s_{ij}-\hat{s}_{ij} \cdot s_{ij}\right)  \right]  }  
    \label{eq8}
\end{align}

\noindent where $b_{ij}$ and $\hat{b}_{ij}$ denote the generated BDM and ideal BDM on position $(i, j)$. $\mathbb{1}[x]= 1$ if x is true and 0 otherwise. $\lambda$ is the threshold to ignore the part with lower loss. $s_{ij}$, $\hat{s}_{ij}$ and $w_{ij}$ represent the prediction, ground truth and weight, respectively.

\subsection{Overall Architecture}
Based on the proposed BDGM and BDGD, we introduce an overall architecture called BDG-Net. We adopt the EfficientNet-B5\cite{tan2019efficientnet} as our backbones to extract multi-stage feature, which is passed to each stage of BDGD as complementary spatial information to guide the polyp segmentation. The BDGM takes the highest three stages of encoder features as input and generates the BDM. We use a U-shaped structure similar to U-Net\cite{ronneberger2015u} with the skip path to pass the encoder features to the decoder.

For the BDGD, we used BDGD-A at the highest two stages, while BDGD-B is used at the lower stages. The output of BDGD is followed by a convolutional layer and a bilinear interpolation upsampling by 2 to generate the segmentation map.

\section{Experimental Results}

\subsection{Data and Metric}
Following PraNet\cite{fan2020pranet}, we conduct experiments on ﬁve challenging polyps datasets: ETIS\cite{silva2014toward}, CVC-ClinicDB\cite{bernal2015wm}, CVC-ColonDB\cite{tajbakhsh2015automated}, CVC300\cite{vazquez2017benchmark} and Kvasir\cite{jha2020kvasir}. Our training set contains 900 randomly selected images in Kvasir and 550 selected images in CVC-ClinicDB, while the remaining 100 pieces of Kvasir and 62 pieces of CVC-ClinicDB are used as test sets. In order to verify the generalization ability of our model, we use three other polyp segmentation datasets as tests, including CVC-ColonDB, ETIS, and CVC300.
The adopted metrics are consistent with PraNet, including mean Dice, mean IoU, weighted F-measure($F^\omega_\beta$) \cite{margolin2014evaluate}, S-measure($S_\alpha$) \cite{fan2017structure}, E-measure($E_\phi^{max}$) \cite{fan2018enhanced} and mean absolute error(MAE) \cite{perazzi2012saliency}. 

\subsection{Comparison with other Method}
Here, we compare the proposed approach with previous state-of-the-art methods, including U-Net\cite{ronneberger2015u}, U-Net++\cite{zhou2018unet++}, ResUNet-mod\cite{zhang2018road}, ResUNet++\cite{jha2019resunet++}, SFA\cite{fang2019selective}, and PraNet\cite{fan2020pranet}. 

\begin{figure} [ht]
\begin{center}
\begin{tabular}{c} 
\includegraphics[height=10cm]{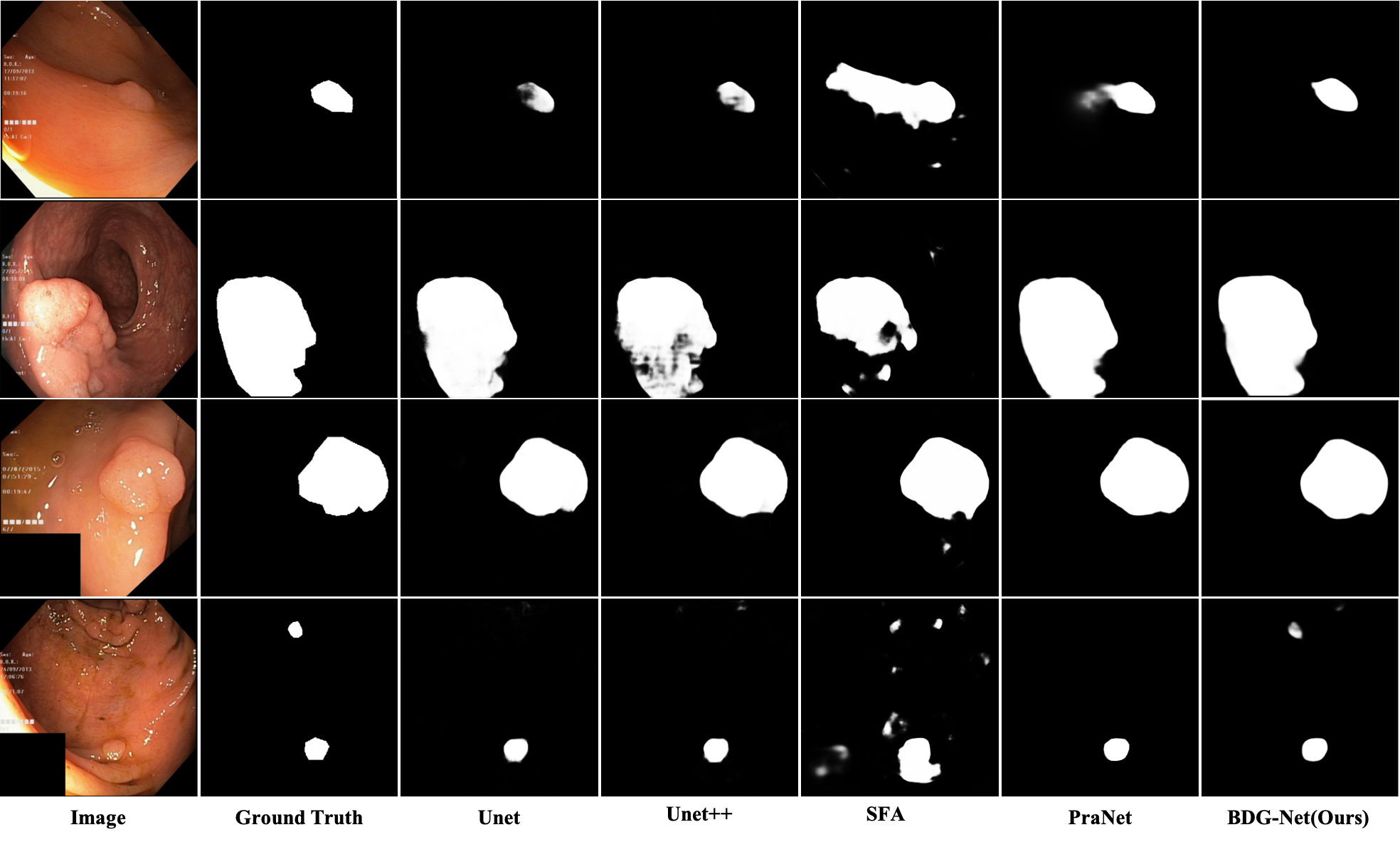}
\end{tabular}
\end{center}
\caption[example]
{ \label{fig6} 
Visual results on Kavsir dataset. (Best viewed in color)}
\hspace{2cm}
\end{figure}

As shown in Table \ref{tab1}, our model obtains mean-Dice of 0.915 and mean-IoU of 0.865 on Kavsir dataset. Other metrics are also better than previous methods remarkably, which shows that our model outperforms all previous state-of-the-art methods. In addition, we calculate the FLOPs under the 352$\times$352 input resolution, and our model achieves the lowest FLOPs, which indicates that our model has the fastest inference speed and lowest computational complexity. As shown in Table \ref{tab2}, our model outperforms all other methods on the CVC-ClinicDB dataset as well. The visual results on Figure \ref{fig6} also show the superiority of our BDG-Net over previous methods. Our method improves the polyps segmentation significantly, especially on smaller objects and unclear boundaries.

\begin{table}[ht]
\caption{Quantitative evaluations on Kavsir dataset. FLOPs is calculated under 352x352 resolution. (`n/a' means data is not available)}
\label{tab1}
\begin{center}
\begin{tabular}{llllllll}
\hline
\rule[-1ex]{0pt}{3.5ex} Methods & mean Dice & mean IoU & $F^\omega_\beta$ & $S_\alpha$ & $E_\phi^{max}$ & MAE & FLOPs(G)\\ 
\hline
\rule[-1ex]{0pt}{3.5ex} U-Net\cite{ronneberger2015u} (MICCAI’15) & \makecell[c]{0.818} & \makecell[c]{0.746} & \makecell[c]{0.794} & \makecell[c]{0.858} & \makecell[c]{0.893} & \makecell[c]{0.055} & \makecell[c]{25.965}\\
\rule[-1ex]{0pt}{3.5ex} U-Net++\cite{zhou2018unet++} (TMI’19) & \makecell[c]{0.821} & \makecell[c]{0.743} & \makecell[c]{0.808} & \makecell[c]{0.862} & \makecell[c]{0.910} & \makecell[c]{0.048} & \makecell[c]{65.925}\\
\rule[-1ex]{0pt}{3.5ex} ResUNet-mod\cite{zhang2018road} & \makecell[c]{0.791} & \makecell[c]{n/a} & \makecell[c]{n/a} & \makecell[c]{n/a} & \makecell[c]{n/a} & \makecell[c]{n/a} & \makecell[c]{n/a}\\
\rule[-1ex]{0pt}{3.5ex} ResUNet++\cite{jha2019resunet++} & \makecell[c]{0.813} & \makecell[c]{0.793} & \makecell[c]{n/a} & \makecell[c]{n/a} & \makecell[c]{n/a} & \makecell[c]{n/a} & \makecell[c]{134.109}\\ 
\rule[-1ex]{0pt}{3.5ex} SFA\cite{fang2019selective} (MICCAI’19) & \makecell[c]{0.723} & \makecell[c]{0.611} & \makecell[c]{0.670} & \makecell[c]{0.782} & \makecell[c]{0.849} & \makecell[c]{0.075} & \makecell[c]{n/a}\\ 
\rule[-1ex]{0pt}{3.5ex} PraNet\cite{fan2020pranet} (MICCAI’20) & \makecell[c]{0.898} & \makecell[c]{0.840} & \makecell[c]{0.885} & \makecell[c]{0.915} & \makecell[c]{0.948} & \makecell[c]{0.030} & \makecell[c]{13.078}\\ 
\hline
\rule[-1ex]{0pt}{3.5ex} BDG-Net (Ours) & \makecell[c]{\textbf{0.915}} & \makecell[c]{\textbf{0.865}} & \makecell[c]{\textbf{0.906}} & \makecell[c]{\textbf{0.923}} & \makecell[c]{\textbf{0.972}} & \makecell[c]{\textbf{0.021}} & \makecell[c]{\textbf{10.840}}\\
\hline
\end{tabular}
\end{center}
\end{table}

To verify the generalization ability of our model, we test on three unseen datasets, including CVC-ColonDB, ETIS, and CVC300. The experimental results show that the proposed method achieves much higher performance than other methods, which reveals the excellent generalization ability of our model.

\begin{table}[ht]
\caption{Quantitative evaluations on other polyp segmentation datasets.} 
\label{tab2}
\begin{center}
\begin{tabular}{lllll}
\hline
\rule[-1ex]{0pt}{3.5ex} Methods & \makecell[c]{CVC-ClinicDB}& \makecell[c]{ColonDB} & \makecell[c]{ETIS} & \makecell[c]{CVC300}\\

\rule[-1ex]{0pt}{3.5ex}  & mDice mIoU & mDice mIoU & mDice mIoU & mDice mIoU \\ 
\hline
\rule[-1ex]{0pt}{3.5ex} U-Net\cite{ronneberger2015u} (MICCAI’15) & \makecell[c]{0.823} \makecell[c]{0.755} & \makecell[c]{0.512} \makecell[c]{0.444} & \makecell[c]{0.398} \makecell[c]{0.335} & \makecell[c]{0.710} \makecell[c]{0.627}\\
\rule[-1ex]{0pt}{3.5ex} U-Net++\cite{zhou2018unet++} (TMI’19) & \makecell[c]{0.794} \makecell[c]{0.729} & \makecell[c]{0.483} \makecell[c]{0.410} & \makecell[c]{0.401} \makecell[c]{0.344} & \makecell[c]{0.707} \makecell[c]{0.624}\\
\rule[-1ex]{0pt}{3.5ex} SFA\cite{fang2019selective} (MICCAI’19) & \makecell[c]{0.700} \makecell[c]{0.607} & \makecell[c]{0.469} \makecell[c]{0.347} & \makecell[c]{0.297} \makecell[c]{0.217} & \makecell[c]{0.467} \makecell[c]{0.329}\\ 
\rule[-1ex]{0pt}{3.5ex} PraNet\cite{fan2020pranet} (MICCAI’20) & \makecell[c]{0.899} \makecell[c]{0.849} & \makecell[c]{0.709} \makecell[c]{0.640} & \makecell[c]{0.628} \makecell[c]{0.567} & \makecell[c]{0.871} \makecell[c]{0.797}\\
\hline
\rule[-1ex]{0pt}{3.5ex} BDG-Net (Ours) & \makecell[c]{\textbf{0.916}} \makecell[c]{\textbf{0.864}} & \makecell[c]{\textbf{0.804}} \makecell[c]{\textbf{0.725}} & \makecell[c]{\textbf{0.756}} \makecell[c]{\textbf{0.679}} & \makecell[c]{\textbf{0.899}} \makecell[c]{\textbf{0.831}}\\ 
\hline
\end{tabular}
\end{center}
\end{table}

\subsection{Ablation Studies}
In this section, We conduct exhaustive ablation study to evaluate the contribution of each component for our BDG-Net. We also investigate the number of skip path and the selection of $\sigma$.

\subsubsection{Effect of BDGM and BDGD}
From Table \ref{tab4}, we can observe that Row No.2 (Backbone + BDGM) outperforms Row No.1 (Backbone), where mean Dice has increased by 1.1\%. Besides, We can also observe that the performance of Row No.3 is significantly better than that of Row No.1 (backbone), increasing the mean Dice and mean IoU by 2.6\%, 3.9\% respectively. To study whether the combination of BDGM and BDGD contributes, Row No.4 (Backbone + BDGM + BDGD) shows the highest performance towards all other settings (No.1 $\sim$ No.3), which improves mean Dice and mean IoU to 0.915 and 0.865, respectively.

\begin{table}[ht]
\caption{Ablation of BDGM and BDGD on Kavsir dataset.} 
\label{tab4}
\begin{center}       
\begin{tabular}{lllllll}
\hline
\rule[-1ex]{0pt}{3.5ex} Setting & \makecell[c]{mean Dice} & \makecell[c]{mean IoU}\\ 
\hline
\rule[-1ex]{0pt}{3.5ex} Backbone & \makecell[c]{0.859} & \makecell[c]{0.780} \\
\rule[-1ex]{0pt}{3.5ex} Backbone + BDGM & \makecell[c]{0.870} & \makecell[c]{0.796} \\
\rule[-1ex]{0pt}{3.5ex} Backbone + BDGD & \makecell[c]{0.896} & \makecell[c]{0.835} \\
\rule[-1ex]{0pt}{3.5ex} Backbone + BDGM + BDGD & \makecell[c]{\textbf{0.915}} & \makecell[c]{\textbf{0.865}} \\ 
\hline
\end{tabular}
\end{center}
\end{table}



\subsubsection{Number of Skip Path and the Choice of $\sigma$}
As shown in Table \ref{tab5}, we can achieve the best result when only using the third and fourth levels of encoder features. Since low-level features are primarily spatial features with much noise, fusing with the first and second-level of encoder features does not bring any gain. We also test the selection of $\sigma$. The experimental results show that when the skip path is employed at the third and fourth stages, the $\sigma$ of 5 can achieve the best results.

\begin{table}[ht]
\caption{Ablation of the number of skip path and $\sigma$ on Kavsir dataset.} 
\label{tab5}
\begin{center}       
\begin{tabular}{lllll}
\hline
\rule[-1ex]{0pt}{3.5ex} Feature Level & $\sigma=1$ & $\sigma=3$ & $\sigma=5$ & $\sigma=7$\\ 
\hline
\rule[-1ex]{0pt}{3.5ex} 4 & 0.911 & 0.912 & 0.911 & 0.906\\
\rule[-1ex]{0pt}{3.5ex} 3, 4 & 0.906 & 0.910 & \textbf{0.915} & 0.911\\
\rule[-1ex]{0pt}{3.5ex} 2, 3, 4 & 0.908 & 0.909 & 0.908 & 0.910\\
\rule[-1ex]{0pt}{3.5ex} 1, 2, 3, 4 & 0.912 & 0.906  & 0.903  & 0.906 \\ 
\hline
\end{tabular}
\end{center}
\end{table}

\section{Conclusions}
In this paper, we present a novel architecture, BDG-Net, for accurate polyp segmentation from colonoscopy images. BDG-Net consists of two novel modules, namely BDGM and BDGD. BDGM generates BDM to estimate the boundary distribution of polyps. Then, the generated BDM is passed to BDGD to guide the polyp segmentation. Extensive experiments demonstrate that the BDG-Net consistently outperforms all state-of-the-art methods across ﬁve challenging datasets on six metrics while maintaining the lowest computational complexity.

\acknowledgments
This work was supported in part by the National Natural Science Foundation of China under Grant 62071086 and Sichuan Science and Technology Program under Grant 2021YFG0296.

\bibliography{report} 
\bibliographystyle{spiebib} 

\end{document}